\documentclass[sigconf]{acmart}

\usepackage{graphicx}
\usepackage{url}
\usepackage{color}
\usepackage{hyperref}
\usepackage{balance}

\setcopyright{none}
\renewcommand\footnotetextcopyrightpermission[1]{} 
\begin{document}

\title{SpaceTEE: Secure and Tamper-Proof Computing\\ in Space using CubeSats}

\author{Yan Michalevsky}
\affiliation{Stanford University}
\email{yanm2@cs.stanford.edu}

\author{Yonatan Winetraub}
\affiliation{Stanford University, SpaceIL}
\email{yonatanw1@stanford.edu}

\begin{abstract}
Sensitive computation often has to be performed in a trusted execution environment~(TEE), which, in turn, requires tamper-proof hardware. 
If the computational fabric can be tampered with, we may no longer be able to trust the correctness of the computation.
We study the idea of using computational platforms in space as a means to protect data from adversarial physical access.
In this paper, we propose SpaceTEE -- a practical implementation of this approach using low-cost nano-satellites called CubeSats.
We study the constraints of such a platform, the cost of deployment, and discuss possible applications under those constraints. 
As a case study, we design a hardware security module solution (called SpaceHSM) and describe how it can be used to implement a root-of-trust for a certificate authority (CA).
\end{abstract}


\maketitle

\sloppy

\section{Introduction}
Extremely sensitive computational operations, involving highly secret data, require running on a computational platform that can be trusted to maintain secrecy and computational integrity.
A common requirement from such a platform is to be tamper-proof, meaning an adversary must not be able to change the way it works. This property should be maintained across the stack, spanning software and hardware. 
Unless complicated and computationally heavy cryptographic techniques from verifiable computation are used (see for example \cite{cryptoeprint:2013:507,cryptoeprint:2015:1243}), the software have to eventually trust the hardware to perform operations correctly. In addition, we would like to ensure that the inner state of the computer is not visible to an adversary.

Assuming secure software implementation, and a functionally correct hardware design, we must prevent post-manufacturing adversarial modifications to the hardware, as well as physical access to its components in order to read data from it.

Tamper-proof hardware is used in various products. One notable example is Hardware Security Modules (HSM).

\subsection{Hardware Security Modules}
HSMs are used for generating and storing cryptographic keys, as well as providing a restricted interface for performing well-defined cryptographic operations, such as message signing and verification.

In Public-Key Infrastructures (PKI), HSMs are used by Certificate Authorities (CA) to generate the signing keys, and to sign other certificates. Thus, the private key never leaves the HSM hardware, and is protected from software attacks on the CA infrastructure.

However, via physical access to the HSM it may potentially be possible to obtain its secrets. HSM vendors such as Safenet and Thales incorporate many precautions and preventative measures to secure their hardware and make it tamper-proof. One defensive measure is to protect them with sensors that identify an attempt to open the HSM enclosure in order to access the board. The FIPS 140-2 US government standard~\cite{fips140-2} specifies the requirements for cryptographic modules that protect sensitive (but unclassified) information for commercial uses.

FIPS 140-2 specifies 4 levels of security. The specification for Level 4 states
\begin{quote}
"Physical security mechanisms provide a complete envelope of protection around the cryptographic module with the intent of detecting and responding to all unauthorized attempts at physical access." \end{quote} as well as \begin{quote}
"protects a cryptographic module against a security compromise due to environmental conditions or fluctuations outside of the module's normal operating ranges for voltage and temperature."
\end{quote} 
The National Institute of Standard and Technology (NIST), Cryptographic Module Validation Program (CMVP) periodically publishes a Validated FIPS 140-1 and FIPS 140-2 Cryptographic Modules list \cite{CMVP}. 
Out of 332, only 13 comply with Level 4 physical security requirements. We were unable to find public pricing and specifications for Level 4 devices, however Level 3 devices in the list cost up to \$50,000 and support 60 - 1200 RSA-2048 signatures per second.

Multiple works on side-channel attacks show that extraction of internal state is possible even without opening the enclosure of a computer~\cite{genkin2014get}, and even without direct contact with it~\cite{genkin2015stealing,genkin2016ecdsa,cryptoeprint:2013:857}, from a distance of up-to tens of meters.

\medskip
In the following we propose using nano-satellites as a trusted, isolated environment for secure computation.

\subsection{CubeSats}
\begin{figure}
    \centering
    \includegraphics[width=.35\textwidth]{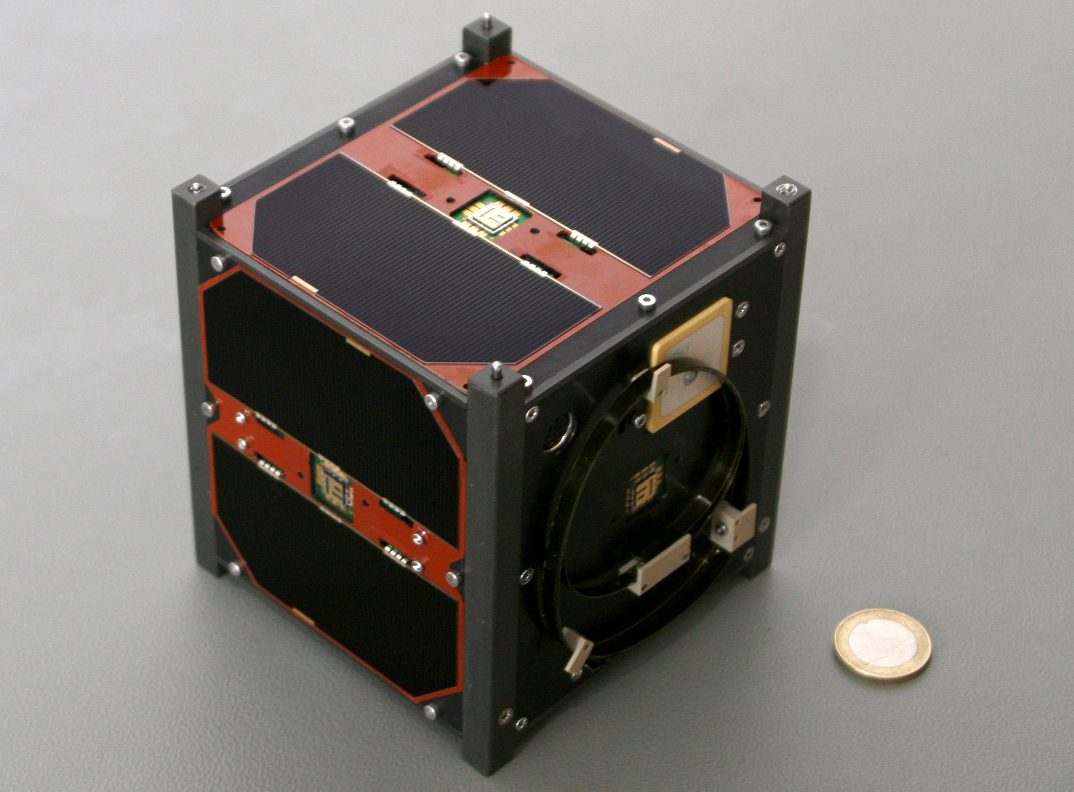}
    \caption{LibreCube: an open-source CubeSat design.}
    \label{fig:libre-cube}
\end{figure}

The CubeSat reference design was proposed in 1999 \cite{helvajian2008small} as a cheap commercial off-the-shelf alternative to traditional expensive satellites. 
The size of a CubeSat 1U is $10\times 10\times 10$~cm, and it weights less than ~1.33~kg.
As of June 24th 2017, 699 CubeSats have been launched to orbit \cite{LaunchedCubesats}, with many active companies in the field generating a total revenue of ~\$2B in 2014 \cite{SatMarket}. CubeSats are mostly used for academic research, however, many disruptive technologies emerged around them recently. 
Increased production volume, and reduction in the cost of launching to orbit, have been driving the cost of launching a CubeSat down in the recent years.
Furthermore, many open-source CubeSat designs exist today (see figure~\ref{fig:libre-cube}), opening the possibilities for innovative solutions that only a decade ago were not available.

\subsection{Trusted Execution in Space} \label{sec:tee-in-space}
Currently, objects in the outer-space are hard to access physically\footnote{Although we can imagine a different situation in the future with advances in mobility in space.}.
However, they can communicate with one another, as well as with terrestrial base-stations. 
Outer-space can therefore serve as a Trusted Execution Environment (TEE), providing strong isolation guarantees, that as of today are very unlikely to be violated even by state-level adversaries.
While anti-satellite weapons have been successfully tested~\cite{antisatellite}, they are currently able, at most, to destroy a satellite, but not to capture it. In a security jargon, destruction constitutes a denial-of-service (DoS) attack.

\subsubsection{Physical Protection and Isolation in Space} \label{sec:physical-protection}
It is very expensive to physically access a satellite launched into space.
Although launching a satellite into space is inexpensive, a rendezvous with one in order to access its hardware is still a difficult mission that requires great expertise and high-end equipment. However, as the cost of space exploration drops, this capability may become more accessible. 
However, even if we will not be able to completely prevent physical access, we can know that such access was attempted and revoke that particular satellite as a reliable root-of-trust.
The North American Aerospace Defense Command (NORAD~\cite{NORAD}), since 1958, has been continuously mapping near-space to identify the course of all objects. NORAD uses a network of ground-based radar systems and telescopes to accurately determine and predict the trajectory of objects of all sizes, as small as a baseball. NORAD's data is freely available online. 
A physical protection layer will involve daily monitoring of the NORAD database (and others) to identify objects which approach our satellite.


\subsubsection {Pre-launch Physical Protection} \label{sec:prelaunch-physical-protection}
If the adversary gets access to the satellite prior to launch, it could tamper with the spacecraft's hardware and software. One way to make the solution tamper-evident\footnote{Meaning, it is easy to see that it has been tampered with (which is a common measure in traditional hardware security modules).} is to coat the computer with kapton tape, and measure the satellite's inertial moment prior to launch. The measurement is compared to the observed rotation rate of the satellite once in orbit. It can be measured by tracking the antennas range-rate as the satellite revolves\cite{marini1972test}, or via direct imaging of the satellite (depending on launch configuration). By measuring the rotation rate changes, we can estimate satellite's actual moment of inertia~\cite{ferguson2008orbit} and verify it is identical to the expected one, measured prior to launch.
Tampering with the coated hardware, or adding new one, inevitably changes the moment of inertia. Further analysis and simulations are required to assess the constraints of this method.
Although it is possible to reintroduce the original inertial moment, it is extremely difficult to do within the time window of a launch effort. In addition, during a launch, the vibrations induced on the satellite from the rocket are likely to shake apart and destroy any component that was not properly installed and tested for vibrations. 

\section{Space-HSM}
We propose a design for a system that can serve as a certificate authority (CA). 
It builds on traditional PKI, combined with the more recent concept of Certificate Transparency~\cite{Laurie2014}. 
We use certificate transparency logs to prevent a powerful attacker, who gains access to the communication channel with the CubeSat, from obtaining forged certificates, similar to the infamous DigiNotar incident~\cite{diginotar}.

\subsection{System Definition}
\begin{figure}
    \centering
    \includegraphics[width=.45\textwidth]{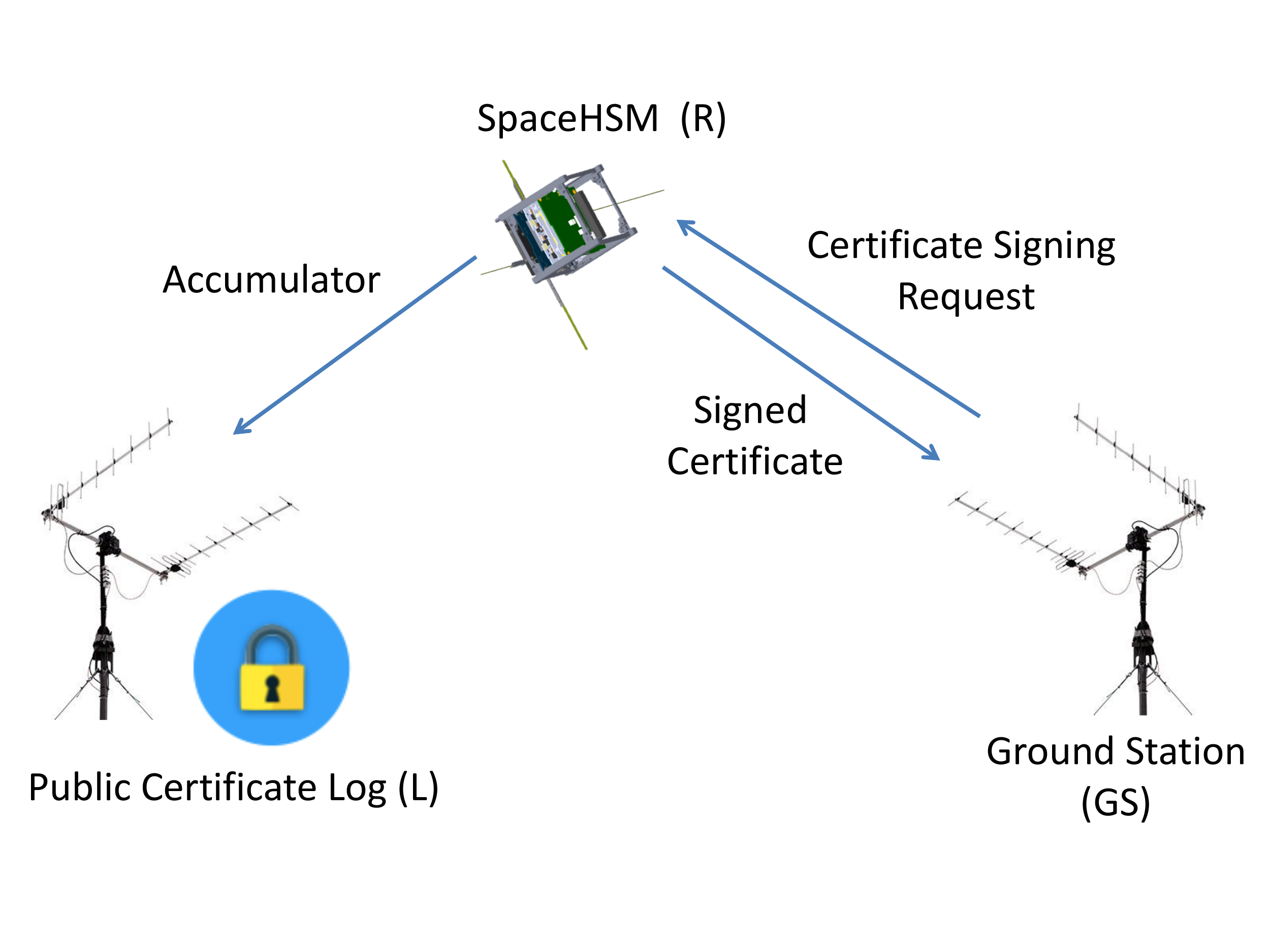}
    \caption{The SpaceHSM System.}
    \label{fig:SysOverview}
\end{figure}

\smallskip\noindent Our system consists of the following entities (see Figure~\ref{fig:SysOverview}): 
\begin{enumerate}
    \item A root-of-trust $\mathcal{R}$ provided by the SpaceHSM satellite. The SpaceHSM handles certificate signing requests, and constantly updates a cryptographic accumulator -- a one-way membership function that enables proving inclusion in an append-only log, and, in our case, represents the complete history of signed requests. 
    The accumulator value is constantly transmitted as the satellite orbits.
    
    \item A terrestrial ground station $\mathcal{G}$ that authenticates to the satellite and delegates requests for certificate signing.
    
    \item A public append-only certificate log $\mathcal{L}$ (similar to Certificate Transparency~\cite{Laurie2014}) that includes the history of all signed certificates. We expect to have a match between the accumulator published by the SpaceHSM and the one computed over the certificate log.
    
    \item A verifier $\mathcal{V}$ that is presented with a certificate, and checks whether it is valid.
    
\end{enumerate}

\subsection{Threat model (without DoS protection)}
As a first step, we state a simplified threat model that demonstrates security but does not consider denial-of-service attacks in the presence of adversaries that can communicate with the satellite.

First and foremost, we assume that our space-based hardware is tamper-proof and the access to it is restricted to a well-defined satellite communication channel\footnote{Satellites often support a command that dumps and transmits all software and data to ground station. We assume no such command is supported in ours.}. 
No physical access to the satellite is possible, as we argue in section \ref{sec:physical-protection}.
We consider a powerful adversary that is occasionally able to gain access to all terrestrial infrastructure, and circumvent all physical security solutions such as locally installed traditional HSM machines, etc. 
It can temporarily obtain wide control of servers connected to the network.
The attacker gains temporary access to the communication channel with the SpaceHSM satellite, which enables her to issue requests to sign forged certificates.

Moreover, we assume that the attacker (and everyone else) has full visibility of the software running on the SpaceHSM, and full read-access to code and data prior to the satellite launch.
Finally, we assume that spoofing transmission from a particular satellite over a considerable period of time, and at many geographic locations (ground stations) is hard\footnote{Essentially it would require a satellite with a similar orbit. However, NORAD would alert about any such existing satellite.}.

\subsection{Protocol description} \label{sec:protocol-no-dos}
\noindent\textbf{Bootstrap.}
Once launched, the SpaceHSM satellite generates a private/public key-pair, and starts broadcasting the public key. 
Once enough ground stations agree on the transmitted public key, we consider the system online. 
Consensus between ground stations can be reached via Byzantine fault-tolerant protocols (see for instance \cite{castro1999practical}), or by relying on a public-key infrastructure (PKI). 
Security of this stage follows from our assumption regarding the difficulty of spoofing communication from the satellite over time and location.

\smallskip\noindent\emph{Boosting trust.} 
An extension of this proposal considers multiple SpaceHSM satellites on different orbits. When two satellites are in line-of-sight of one another, they can exchange attestations of their respective public keys. Then, each one of them can broadcast the signed attestation, increasing the trust in the public key identity.

\smallskip\noindent\textbf{Certificate Request.}
A request to sign a certificate is transmitted to the SpaceHSM, encrypted under its public key for privacy.
The SpaceHSM generates a signed certificate, and updates an accumulator representing the history of all signature requests. 
We use a constant-size accumulator, which suits well the present bandwidth and storage constraints. 
A Merkle hash-tree~\cite{merkle1989certified}, or an RSA accumulator~\cite{Benaloh1994}, can serve the purpose.

It is important that the certificate is actually signed, in addition to updating the accumulator since it guarantees to anyone presented with the signed certificate, that the SpaceHSM updated its accumulator. That, in turn, guarantees that any inconsistency would have been noticed by monitoring parties.

\smallskip\noindent\textbf{Certificate Verification.}
Anyone presented with the signed certificate can verify its authenticity using the public key broadcast at the bootstrap stage. 

\smallskip\noindent\textbf{Certificate Log Update.}
We expect every signed certificate to be submitted to the publicly accessible certificate log. When log server $\mathcal{L}$ receives a new signed certificate, it verifies the signature using $\mathcal{R}$'s public key, and appends it to the log. Anyone can then compute an accumulator over the log and compare it to the one being broadcast by the SpaceHSM $\mathcal{R}$. We call such parties monitors.

\subsubsection{Security}
Security and trust stem from correspondence between the accumulator value transmitted by the SpaceHSM to the value that is can be computed over the publicly readable terrestrial certificate log. An adversary may be able to take over a ground station and request to sign a forged certificate that is not subsequently submitted to the public terrestrial certificates log. However, in that case, the SpaceHSM's accumulator is updated to a new value, resulting in a mismatch that would be readily noticed by the log monitors.

\subsection{Threat model (with DoS)}
This extended threat model considers Denial-of-Service (DoS) attacks on the system.
The protocol in section \ref{sec:protocol-no-dos} enables anyone who can communicate with the satellite to request the SpaceHSM to sign a certificate. Failure to submit the signed certificate to the log results in an inconsistency between the publicly computeable accumulator and the one being broadcast by the SpaceHSM.
Such a mismatch results in a service disruption. Without appending the adversary's certificate to the log we cannot synchronize it with the state of the SpaceHSM. We therefore need a reset procedure for the SpaceHSM.

We assume that the ground station and the satellite establish a secure communication channel, using a shared secret key.
The adversary is occasionally able to steal this shared key, which results in a disruption to the service.
In this threat model, the adversary has no access to the SpaceHSM prior to the launch. Finally we assume the existence of a secure offline storage on earth.

\medskip
\subsection{Protection against adversarial requests}
This measure addresses the case of an adversary that attempts to cause denial of service by frequently issuing certificate signing requests and transmitting them to the satellite.
We stress that even if such an incident occurs, it has no effect on the trustworthiness of the certificates signed by the SpaceHSM. This procedure is only for the sake of preventing DoS attacks by parties that are normally not authorized to send certificate signing requests directly.

One straightforward solution is to have the SpaceHSM broadcast every signed certificate, such that it is visible to all certificate log servers, instead of returning it via the private channel accessible only to the ground-station. In the following, we suggest another alternative.

We propose a reset procedure when a mismatch between the satellite broadcast and the active log is encountered.
A random seed is generated prior to launch and a cryptographic pseudo-random number generator (PRG) is seeded. The state of the PRG is programmed
intro the SpaceHSM, and is used to derive the first shared symmetric key. 
This state is also stored in a secure and inaccessible offline terrestrial storage.
The offline storage prevents any online attacker from accessing the seed.
When an inconsistency is noticed, we fetch the stored PRG state and use it to generate the next key.
If the SpaceHSM fails to decrypt the channel using the $i$-th (current) key, it generates the $i+1$ key and attempts decryption, in order to automatically identify the state transition. The previous certificate log is no longer extended with new certificate and its state is frozen. 
Instead, $\mathcal{L}$ initializes a new log that is extended with new certificates signed after the key transition. 
The old log is still accessible, and enables verifying inclusion of certificates signed prior to the key transition.

\section{Implementation} \label{sec:implementation}

\subsection{Space Segment Hardware} \label{sec:hardware}
\begin{figure}
    \centering
    \includegraphics[width=.35\textwidth]{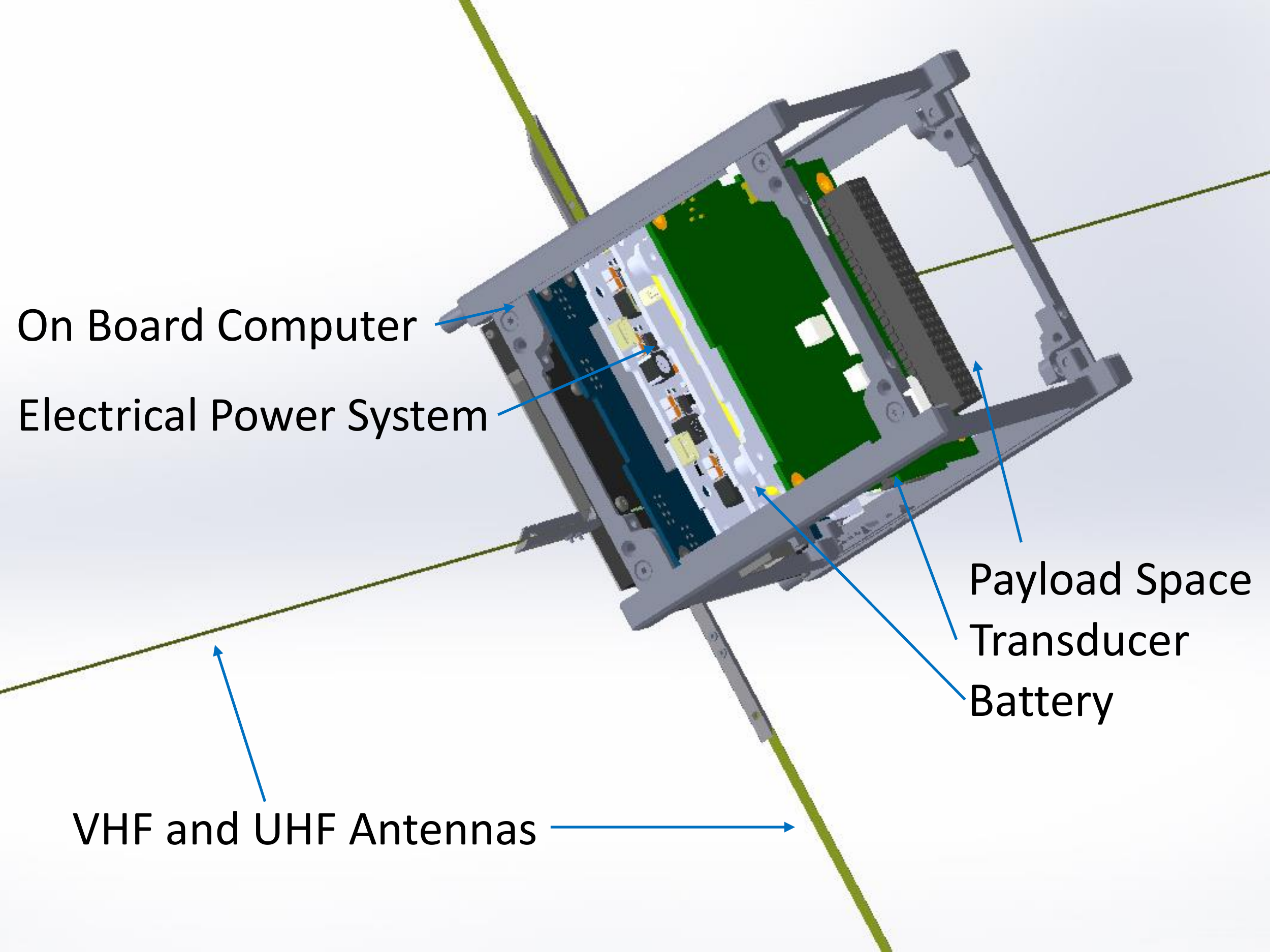}
    \caption{Proposed satellite CAD model (solar panels and secondary structure are not shown).}
    \label{fig:HSMCAD}
\end{figure}

We propose to use commercial off-the-shelf CubeSat parts that were tested in space. 
The satellite is launched to a Sun Synchronous Orbit (SSO), which is a common design choice for CubeSats. 
Table~\ref{tab:bom} provides a bill of materials, and Figure~\ref{fig:HSMCAD} provides a candidate CAD design.

\begin{table}
    \centering
    \begin{tabular}{|c|c|c|} 
     \hline
     Component & Model Name & Notes \\ 
     \hline
     Structure & 1U CubeSat  &   \\ 
     OBC  & Cube Computer \cite{OBCDataSheet} &  ARM Cortex-M3 \\  
     Power system &  CS 1U Bundle A & 10~Whr battery \\
     Solar panels & Azurspace panels 
     \cite{AzurSpaceDataSheet} & On all 6 sides \\
     Transceiver & ISIS VHF/UHF \cite{VHFUHFDataSheet} & Full-duplex \\
     Antennas    & ISIS dipole  & \\
     \hline
    \end{tabular}
    \caption{Bill of materials.}
    \label{tab:bom}
\end{table}

The main on-board computer (OBC) is a Cube Computer~\cite{OBCDataSheet}, consisting of a 32-bit ARM Cortex-M3 with 4MB Flash memory for code storage, and $2\times 1$MB SRAM memory for data storage. In our design, we connect it to an external 48MHz clock. 
The Cube Computer underwent radiation testing and was qualified for space missions. 
Its error detection and correction (EDAC) hardware is able to detect and correct Single Event Upset (SEU) and Single Event Latch (SEL).
By referring to published pricings of the components in Table \ref{tab:bom}, we estimate that the marginal cost for building and launching a SpaceHSM satellite to polar low Earth orbit is on the order of 170,000 USD, or about 95,000 USD per HSM, if two are bundled together into one satellite . It is higher, but comparable to prices of other HSMs on the market that provide FIPS 140-2 Level 4 physical security. 
As we can see in Figure \ref{fig:HSMCAD}, we have enough space to accommodate 2 independent (except for power supply) HSMs, each with its own antenna, transceiver and computer in one CubeSat.

\subsection{Power Budget}
In a typical SSO orbit, satellites have ~45 minutes of daylight and 45 minutes of night-time, where the satellite is in the Earth's shadow (eclipse).

\bigskip\noindent\textbf{Daylight Input Power Generation.}
The satellite will be powered using 6 high-end solar panels with 30\% efficiency \cite{AzurSpaceDataSheet}. The total average power input is  computed (similarly to \cite{arnold2012energy}) to be $3.82[W]$.

\smallskip\noindent\textbf{Daytime Power Consumption.} 
Three components contribute to the daytime power consumption: the on-board computer, with an average of $0.2[W]$~\cite{OBCDataSheet};
the transceiver, which consumes $1.7[W]$ during transmission and $0.2[W]$ while receiving~\cite{VHFUHFDataSheet}, and given a duty cycle of 30\% at transmission, results in average power consumption of $0.65[W]$; and finally battery recharging after night operation, which consumes $0.85[W]$.

The total consumption by a single SpaceHSM is $1.7 [W]$, which is 44\% of the generated power, meaning the satellite has enough power to sustain two SpaceHSMs.

\smallskip\noindent\textbf{Night-time Power Consumption.} 
At night, the satellite draws power from a battery to operate both the transceiver and the on-board computer with the same power consumption as in daylight: $0.85[W]$. In 45 min of nighttime, the satellite consumes $0.64[W\cdot hr]$ drawn from the $10[W\cdot hr]$ battery. 
A satellite's depth of discharge is very low (less then 10\%), which guarantees long battery life to support a satellite in orbit for several years \cite{BatLifeSpan}.

\subsection{Link Budget and Bit Rates} \label{sec:link-budget}
We consider a typical certificate signing request size of 2.5~KBytes transmitted to the satellite, and a 256 byte long RSA-2048 signature and a 32 byte long accumulator using a SHA-256 hash, broadcast in response. 
We assume a standard CubeSat datalink using the AX.25 protocol~\cite{AX25}. An AX.25 frame adds a maximum overhead of 35 bytes, that wraps 256 bytes of data (Info field)~\cite{QB50Allocation1}. 5 bytes of the Info field are used as a message header~\cite{QB50Allocation2}. 
We are left with 251 bytes per packet available for custom data.
Therefore, for each certificate request, the SpaceHSM receives 11 AX.25 packets (24,000 bit) and sends one AX.25 packet (600 bit).

We do not elaborate on the link budget since our transducer was successfully tested in SSO, with a standard armature radio ground station. 
According to \cite{VHFUHFDataSheet}, we expect an uplink of 1200bps - enough for one certificate request every 20 seconds. 
The downlink rate of 2400 bps at a 30\% duty cycle enables sending a reply every second. Due to orbital mechanics, we can communicate with the satellite for 10min every 90min using Svalbard ground station.

These rates can be improved in the future by a factor of 1,000 by using an S-Band transceiver capable of uplink rates of 2~Mbps~\cite{SBandTrancever}. 
However, using it, requires a monopole / dipole S-Band antenna and a suitable ground station. 
They can be custom-ordered, but are not yet widely available commercially for CubeSats. 
We believe that this technology will be widely used within a few years, which would enable handling 50 certificate signing requests per second.

\subsection{Software} \label{sec:software}
SpaceHSM requires crypto algorithms, including symmetric encryption (AES-128 and AES-256) and, depending on the chosen PKI algorithms, RSA-2048 and RSA-4096 signatures, ECDSA signatures, and Curve25519.
Due to limited resources we use a small cryptographic library. To give a few examples:
TweetNaCl~\cite{bernstein2014tweetnacl} is a tiny cryptographic library with compiled  binary size of 11 KB, support around 40 RSA-2048 decryptions per second on an Arm Cortex M3.
mbedTLS~\cite{mbedtls}, while larger, is very functional and still fits within our RAM budget. Its compiled binary takes ~125~KB. 
Another alternative is the RTOS ready SharkSSL library, that was specifically benchmarked on ARM Cortex M3 \cite{SharkSSL}. 
Its binary size is approximately 15~KB.
Finally, WolfSSL~\cite{wolfSSL} is a small commercial open-source library that supports TLS 1.3, ChaCha20, Curve25519 in addition to the standard crypto algorithms.

\subsection{Fault protection} \label{sec:fault-protection}
It is important to protect cryptographic protocols from random (or injected) faults~\cite{boneh1997importance}.
While random faults can potentially be caused by radiation in space, the tests mentioned in section~\ref{sec:hardware} show that all bit-flips (SEU) are identified by the EDAC, and are corrected. Effectively, it implies that random faults are very unlikely to occur during the lifespan of the SpaceHSM.
Faults can potentially be injected by heating the satellite using a laser beam.
Nevertheless, we protect against fault-attacks on signature algorithms by verifying the computed signature.

\section{Additional Applications} \label{sec:future-work}
Our proposal has significant advantages over terrestrial computers in the ability to isolate, and to protect against physical access. 
However, its drawback is the relatively low communication bandwidth, as well as the relatively limited processing power, due to usage of micro-processors. 
Notably, the link is not symmetric -- the bottleneck is the uplink to the satellite, which precludes uploading large amounts of data to the satellite for computing in an isolated environment.
In classic satellite applications, the downlink rate is usually the bottleneck, rather than the uplink. 
Therefore most existing products are tailored to have high downlink at the expense of a high uplink rate. 
Further research is needed to better understand how uplink rate can be improved at the expense of a lower downlink rate.
Additional applications and implementations of secure and tamper proof computing in space may include:

\medskip\noindent\textbf{Trusted Public Parameter Generator.}
Cryptographic schemes often rely on a trusted setup, that occurs once, and outputs common public parameters. Any secret values generated during the setup must be ``forgotten" after, and remain inaccessible. The SpaceTEE platform, with its support for random number generation and cryptographic functionality, is a perfectly suitable party for executing such algorithms.

\medskip\noindent\textbf{Trusted Party for Cryptographic Protocols.}
Many cryptographic protocols (electronic voting, privacy-preserving aggregation) require a trusted party, that performs a certain computation without revealing the internal state, and that cannot be corrupted by an adversary. Our CubeSat, loaded with suitable software can potentially be such a party. Currently, it suits only moderate workloads, which can however be useful for certain protocols. A simple e-voting protocol involves submitting inputs encrypted under an additively-homomorphic encryption scheme. The inputs are "summed-up", resulting in an encryption of the aggregate value. A trusted party then decrypts the ciphertext and outputs the result. Such a trusted party can be implemented using SpaceTEE.

\medskip\noindent\textbf{Trusted Mining.}
The SpaceTEE can serve as a trusted miner for cryptocurrencies similar to Bitcoin and Ethereum. A major concern in cryptocurrencies, is that miners with a large percentage of the hashing power can adopt a non-standard mining strategy in an attempt to profit beyond the standard transaction fees~\cite{eyal2014majority,kroll2013economics,courtois2014subversive}. For instance, they can ignore the most recent block, etc. Execution in a trusted environment can guarantee strict adherence to a certain mining strategy, decided prior to launch.

\medskip\noindent\textbf{Trusted Timestamping} enables securely keeping track of creation and modification of documents and media, and can be useful for instance for copyright purposes. A client can request the SpaceHSM to sign a message containing a hash of the document and a timestamp at the time of signing, and later reveal the original document, i.e.~the hash function preimage, claiming ownership.

\section{Related Work}

The project on firstspacebank.com~\cite{spacebank} sets to provide a secure and decentralized satellite banking platform. The website, unfortunately, provides no further details beyond this short statement.
Laurie's technical report on Certificate Transparency~\cite{Laurie2014} provides an excellent explanation of the motivation behind it and its design.
There are various proposals for implementation of cryptographic accumulators, starting with the original paper by Benaloh and de Mare \cite{Benaloh1994}. Tremel~\cite{tremel2013real} provides a review of various accumulators, and compares the performance of their real-world implementations.
The following survey, \cite{CubesatSurvey}, demonstrates new and emerging applications for CubeSats.

\section{Conclusion}
In this work, we present a novel idea of using computational platforms in space as a means to protect data from adversarial physical access. We demonstrated one application of such tamper-proof platform by showing how Hardware Security Modules for a PKI ecosystem can be implemented using an inexpensive CubeSat that can be built using readily available commercial off-the-shelf hardware.
A significant impediment to using space-based tamper-proof computation today, is the limited communication bandwidth to and from the satellite, as well as the relatively low processing power. However, in the near future significant advancements can be done in the field.

\section*{Acknowledgements}
We would like to thank Elad Sagi and Avi Barliya for reviewing the satellite concept and design, and Henry Corrigan-Gibbs for reviewing and commenting on the security aspects of the system. We thank the anonymous reviewers for providing valuable comments on the paper.
Finally, we thank LibreCube for licensing the artistic materials on their website under a Creative Commons license.

\balance
\bibliographystyle{abbrv}
\bibliography{main}

\end{document}